\documentclass[
,final            
]{aipproc}
\layoutstyle{8x11single}
\usepackage{graphicx}
\graphicspath{{fig/}}
\usepackage{bm}
\usepackage{amsmath}
\usepackage{amssymb}
\usepackage{color}
\begin{document}
\title{Time Dependent Ginzburg-Landau Equation for Sheared Granular Flow}
\author{Kuniyasu Saitoh}
{address={Faculty of Engineering Technology, University of Twente, Enschede, the Netherlands}}
\author{Hisao Hayakawa}
{address={Yukawa Institute for Theoretical Physics, Kyoto University, Sakyo-ku, Kyoto, Japan}}
\begin{abstract}
The time dependent Ginzburg-Landau equation for a two-dimensional granular shear flow
is numerically solved, where we study both the transient dynamics and the steady state
of the order parameter. The structural changes of the numerical solutions are qualitatively
similar to the shear bands observed in the discrete element method (DEM) simulation of
the two-dimensional granular shear flow.
\end{abstract}
\maketitle
\section{Introduction}
Flows of granular particles have been well studied due to the importance in technology,
engineering, geophysics, astrophysics, applied mathematics and physics \cite{luding1,luding2,bri,gold}. 
The characteristic properties of granular flows are mainly caused by the inelastic
collisions between particles \cite{jeager}. Among various studies of granular flows, the
study of the granular gases under a plane shear plays an important role from many
aspects, e.g., the application of the kinetic theory to granular gases
\cite{sela,santos,lun,dufty1,dufty2,lutsko1,lutsko2,lutsko3,jr1,jr2},
the shear band formation in moderately dense granular gases \cite{tan,saitoh},
the long-time tail and the long-range correlation
\cite{kumaran1,kumaran2,kumaran3,orpe1,orpe2,rycroft,lutsko0,otsuki0,otsuki1,otsuki2},
the pattern formation in dense granular flow \cite{louge1,louge2,louge3,louge4,khain1,khain2},
the determination of the constitutive equation for dense granular flow \cite{midi,cruz,hatano1},
as well as the jamming transition \cite{hecke,hatano2,hatano3,otsuki3,otsuki4,otsuki5,otsuki6}.

The granular hydrodynamic equations derived by the kinetic theory well describe the
dynamics of moderately dense granular gases \cite{lun,dufty1,dufty2,lutsko1,lutsko2,lutsko3,jr1,jr2},
though the validity of the kinetic theory is questionable in the case of granular gases,
because of the lack of scale separation and the existence of the long range correlations, etc \cite{gold}.
For a granular shear flow, a homogeneous state is unstable in the presence of the plane shear
\cite{linear1,linear2,layer1,layer2,layer3,layer4,layer5}. As a result, two shear bands are
formed near the boundary, and they collide to form one shear band in the center region under
a physical boundary condition \cite{tan,saitoh}. A similar shear band formation is also
observed under the Lees-Edwards boundary condition. It is known that both the transient
dynamics and the steady state of the hydrodynamic fields can be approximately reproduced
by the granular hydrodynamic equations \cite{saitoh}.

To understand the shear band formation after the homogeneous state becomes unstable,
we have to develop a weakly nonlinear analysis. Recently, Shukla and Alam carried out
a weakly nonlinear analysis of granular shear flow, where they derived the
Stuart-Landau equation of the order parameter defined as the amplitude of disturbance
to the hydrodynamic fields under a physical boundary condition starting from a set of
granular hydrodynamic equations \cite{shukla1,shukla2,shukla3}. They found the existence
of subcritical bifurcations
in both relatively dilute and dense regions, while a supercritical bifurcation appears
in the moderate density region. The Stuart-Landau equation, however, does not include
any spatial degrees of freedom and cannot be used to study the time evolution of the
shear band.

It is also notable that Khain found the coexistence of a solid phase and a liquid phase
in the molecular dynamics simulation of a dense granular shear flow \cite{khain1,khain2}.
He also demonstrated the hysteresis of the order parameter defined as the difference of
the densities between the boundary and the center region. It should be noted, however, the
mechanism of the subcritical bifurcation based on a set of hydrodynamic equations differs
from that observed in the jamming transition of frictional particles \cite{otsuki6}.

In our previous work, we have developed the weakly nonlinear analysis of a two-dimensional
granular shear flow and derived the time dependent Ginzburg-Landau (TDGL) equation of the
order parameter defined as the amplitude of disturbance to the hydrodynamic fields under
the Lees-Edwards boundary condition \cite{saitoh1}. We introduced a hybrid approach to the
weakly nonlinear analysis and the resultant TDGL equation is a two-dimensional partial
differential equation associated with the time dependent diffusion coefficients \cite{saitoh1}.
The TDGL equation derived by the hybrid approach is useful to understand the structural
changes of shear bands and we also discussed the bifurcation of the order parameter.
However, we have not analyzed the solution of the TDGL equation and compared the solution
with the DEM simulation yet.

In this paper, we numerically solve the TDGL equation derived in Ref. \cite{saitoh1}
to exhibit the transient dynamics and the steady state of the order parameter.
In the following, we review our previous work of the weakly nonlinear analysis
at first. At second, we present the numerical solution of the TDGL equation.
Finally, we discuss and conclude our results.

\section{Overview of weakly nonlinear analysis}

In this section, we review our previous results of the weakly nonlinear analysis \cite{saitoh1}.
At first, we introduce the hydrodynamic equations of the area fraction, the velocity fields
and the granular temperature. At second, we derive the one-dimensional TDGL equation by the
ordinary weakly nonlinear analysis. At third, we derive the two-dimensional TDGL equation
by adopting the hybrid approach to the weakly nonlinear analysis.

\subsection{Basic Equations}

Let us introduce our setup and basic equations. We adopt the Lees-Edwards boundary condition
for the boundary of a two-dimensional granular shear flow, where the upper and the lower image
cells move to the opposite directions with a constant speed $U/2$, and the distance between the
upper and the lower image cells is given by $L$ \cite{lees}. Since we assume the two-dimensional
granular disks are identical, the mass, the diameter and the restitution coefficient of granular
disks are given by $m$, $d$ and $e$, respectively. In the following argument, we scale the mass, the
length and the time by $m$, $d$ and $2d/U$, respectively. Therefore, the shear rate $U/L$ is
reduced to $\epsilon \equiv 2d/L$ in our units and $\epsilon$ is a small parameter in the
hydrodynamic limit $L\gg d$.

We employ a set of granular hydrodynamic equations derived by Jenkins and Richman
\cite{jr1}. Although their original equations include the angular momentum and the spin
temperature, the spin effects are localized near the boundary \cite{mitarai} and the effect
of rotation can be absorbed in the normal restitution coefficient, if the friction constant
is small \cite{jz,yj}. Therefore, we neglect the rotational degrees of freedom and
the dimensionless hydrodynamic equations are given by
\begin{eqnarray}
\left( \partial_t + \mathbf{v} \cdot \nabla \right)
\nu &=& - \nu \nabla \cdot \mathbf{v} \label{eq:hy1} \\
\nu \left( \partial_t + \mathbf{v} \cdot \nabla \right)
\mathbf{v} &=& - \nabla \cdot \mathsf{P} \label{eq:hy2} \\
\left( \nu / 2 \right) \left( \partial_t + \mathbf{v} \cdot \nabla \right)
\theta &=& - \mathsf{P} : \nabla \mathbf{v} - \nabla \cdot \mathbf{q} - \chi \label{eq:hy4}~,
\end{eqnarray}
where $\nu$, $\mathbf{v}=(u,w)$, $\theta$, $t$ and $\nabla=(\partial/\partial_x,\partial/\partial_y)$
are the area fraction, the dimensionless velocity fields, the dimensionless granular
temperature, the dimensionless time and the dimensionless gradient, respectively.
The pressure tensor $\mathsf{P}=(P_{ij})$, the heat flux  $\mathbf{q}$ and the energy
dissipation rate $\chi$ are given by
\begin{eqnarray}
P_{ij} &=& \left[ p(\nu)\theta - \xi(\nu)\theta^{1/2}
\left( \nabla \cdot \mathbf{v} \right) \right] \delta_{ij} - \eta(\nu)\theta^{1/2} e_{ij}~,
\label{eq:stress} \\
\mathbf{q} &=& - \kappa(\nu) \theta^{1/2} \nabla \theta - \lambda(\nu) \theta^{3/2} \nabla \nu~,
\label{eq:heat} \\
\chi &=& \frac{1-e^2}{4\sqrt{2\pi}}\nu^2 g(\nu) \theta^{1/2} \left[ 4\theta - 3\sqrt{\frac{\pi}{2}}
\theta^{1/2} \left( \nabla \cdot \mathbf{v} \right) \right]~,
\label{eq:chi}
\end{eqnarray}
respectively, where $p(\nu)\theta$, $\xi(\nu)\theta^{1/2}$, $\eta(\nu)\theta^{1/2}$,
$\kappa(\nu)\theta^{1/2}$ and $\lambda(\nu)\theta^{3/2}$ are the dimensionless forms
of the static pressure, the bulk viscosity, the shear viscosity, the heat conductivity
and the coefficient associated with the density gradient, respectively, and
$e_{ij}\equiv(\nabla_jv_i+\nabla_iv_j-\delta_{ij}\nabla\cdot\mathbf{v})/2$~$(i,j = x,y)$
is the deviatoric part of the strain rate. The explicit forms of them are listed in Table
\ref{tab:subfunctions}, where
\begin{equation}
g(\nu) = \frac{1-7\nu/16}{\left( 1-\nu \right)^2} \label{eq:radial}
\end{equation}
is the radial distribution function at contact which is only valid for $\nu < 0.7$
\cite{gnu4,gnu3,gnu2,gnu1}.

A set of homogeneous solutions of Eqs. (\ref{eq:hy1})-(\ref{eq:hy4}) is readily
found as $\phi_0\equiv(\nu_0,\epsilon y,0,\theta_0)$, where $\nu_0$ and $\theta_0\propto\epsilon^2/(1-e^2)$
are the mean area fraction and the mean granular temperature, respectively. In our
analysis, $\theta_0\sim O(1)$. Therefore, $\epsilon\sim\sqrt{1-e^2}$ and the small
$\epsilon$ corresponds to the small inelasticity, where $e$ is close to unity \cite{saitoh1}.
\begin{table}
\begin{tabular}{lll}
\hline
$p(\nu)$ &$=$& $\frac{1}{2}\nu \left[ 1+(1+e)\nu g(\nu) \right]$ \\
$\xi(\nu)$ &$=$& $\frac{1}{\sqrt{2\pi}} (1+e)\nu^2 g(\nu)$ \\
$\eta(\nu)$ &$=$& $\sqrt{\frac{\pi}{2}} \left[ \frac{g(\nu)^{-1}}{7-3e}
+ \frac{(1+e)(3e+1)}{4(7-3e)}\nu + \left( \frac{(1+e)(3e-1)}{8(7-3e)} + \frac{1}{\pi} \right)(1+e)\nu^2 g(\nu) \right]$ \\
$\kappa(\nu)$ &$=$& $\sqrt{2\pi} \left[ \frac{g(\nu)^{-1}}{(1+e)(19-15e)}
+ \frac{3(2e^2+e+1)}{8(19-15e)}\nu + \left( \frac{9(1+e)(2e-1)}{32(19-15e)} + \frac{1}{4\pi} \right)(1+e)\nu^2 g(\nu) \right]$ \\
$\lambda(\nu)$ &$=$& $- \sqrt{\frac{\pi}{2}}\frac{3e(1-e)}{16(19-15e)}\left[ 4(\nu g(\nu))^{-1} + 3(1+e) \right] \frac{d \left( \nu^2g(\nu) \right)}{d\nu}$ \\
\hline
\end{tabular}
\caption{The functions in Eqs.(\ref{eq:stress})-(\ref{eq:chi}).}
\label{tab:subfunctions}
\end{table}
\subsection{Weakly Nonlinear Analysis}

The homogeneous solution $\phi_0$ is linearly unstable and the disturbance to the hydrodynamic
fields $\hat{\phi}$ with the most unstable mode develops as time goes on \cite{linear1,linear2,layer1,layer2,layer3,layer4,layer5}.
To understand the time evolution of $\hat{\phi}$, we need to carry out a weakly nonlinear analysis.
For this purpose, we introduce the long time scale $\tau\equiv\epsilon^2 t$ and the long length scales
$(\xi,\zeta)\equiv\epsilon(x,y)$, respectively. Then, the \textit{neutral solution} is given by
\begin{equation}
\hat{\phi}_\mathrm{n} = A^{\rm L}(\zeta,\tau)\phi^{\rm L}_{q_c}e^{i q_c \zeta}
+ \mathrm{c.c.}~, \label{eq:neutral}
\end{equation}
where $\mathrm{c.c.}$ represents the complex conjugate and $\phi^{\rm L}_{q_c}$ is the Fourier
coefficient of the most unstable mode $\mathbf{q}_c=(0,q_c)$. The amplitude $A^{\rm L}(\zeta,\tau)$
is independent on $\xi$, because any modes in the sheared frame $\mathbf{q}(\tau)=(q_\xi,q_\zeta-\epsilon tq_\xi)$
with $q_\xi\neq 0$ are linearly stable \cite{linear1,linear2,layer1,layer2,layer3,layer4,layer5}.

We expand $A^{\rm L}(\zeta,\tau)$ into the series of $\epsilon$ as
\begin{equation}
A^{\rm L}(\zeta,\tau) = \epsilon A^{\rm L}_1(\zeta,\tau) + \epsilon^2 A^{\rm L}_2(\zeta,\tau)
+ \epsilon^3 A^{\rm L}_3(\zeta,\tau) + \dots \label{eq:amp_exp}
\end{equation}
and substitute Eqs. (\ref{eq:neutral}) and (\ref{eq:amp_exp}) into the hydrodynamic equations
(\ref{eq:hy1})-(\ref{eq:hy4}). Collecting each order terms of $\epsilon$, we find the first
non-trivial equation of $A^{\rm L}_1(\zeta,\tau)$ at $O(\epsilon^3)$, which is the TDGL equation
\begin{equation}
\partial_\tau A^{\rm L}_1 = \sigma_c A^{\rm L}_1 + D \partial_\zeta^2 A^{\rm L}_1
+ \beta A^{\rm L}_1 |A^{\rm L}_1|^2~, \label{eq:GL30}
\end{equation}
where $D$ and $\beta$ are the functions of $\nu_0$ (listed in Table 2 of
Ref. \cite{saitoh1}) and $\sigma_c$ is the maximum growth rate at $\mathbf{q}_c$ scaled by $\epsilon^2$.
Because of the scaling relations $D=\bar{D}$ and $\beta=\epsilon\bar{\beta}$, Eq. (\ref{eq:GL30}) is
rewritten as the equation of the scaled amplitude $\bar{A}^{\rm L}_1(\zeta,\tau)\equiv\epsilon^{1/2}A^{\rm L}_1(\zeta,\tau)$
\begin{equation}
\partial_{\tau} \bar{A}^{\rm L}_1 = \sigma_c \bar{A}^{\rm L}_1
+ \bar{D} \partial_{\zeta}^2 \bar{A}^{\rm L}_1 +
\bar{\beta} \bar{A}^{\rm L}_1 |\bar{A}^{\rm L}_1|^2~.
\label{eq:red_TDGL3_layer}
\end{equation}
The solution of Eq. (\ref{eq:red_TDGL3_layer}) converges only if $\bar{\beta}<0$, i.e.,
in the case of a \textit{supercritical bifurcation}.

Developing a similar procedure till $O(\epsilon^5)$, we obtain the higher order equation
\begin{equation}
\partial_\tau \check{A}^{\rm L} =
\sigma_c \check{A}^{\rm L} + \bar{D} \partial_\zeta^2 \check{A}^{\rm L}
+ \bar{\beta} \check{A}^{\rm L} |\check{A}^{\rm L}|^2
+ \epsilon \bar{\gamma} \check{A}^{\rm L} |\check{A}^{\rm L}|^4 + O(\epsilon^3)~,
\label{eq:GL5_red}
\end{equation}
where $\check{A}^{\rm L}(\zeta,\tau)\equiv\epsilon^{1/2} [A^{\rm L}_1(\zeta,\tau) +
\epsilon A^{\rm L}_2(\zeta,\tau) + \epsilon^2 A^{\rm L}_3(\zeta,\tau)]$ and $\bar{\gamma}$
is the function of $\nu_0$ (listed in Table 2 of Ref. \cite{saitoh1}). If $\bar{\gamma}<0$,
the solution of Eq. (\ref{eq:GL5_red}) converges even if $\bar{\beta}>0$, i.e., in the case
of a \textit{subcritical bifurcation}.

\subsection{Hybrid Approach to the Weakly Nonlinear Analysis}

The amplitudes $\bar{A}^{\rm L}_1(\zeta,\tau)$ and $\check{A}^{\rm L}(\zeta,\tau)$
are independent of $\xi$ and cannot describe the two-dimensional structure of shear bands.
Thus, we need to introduce a new approach to the weakly nonlinear analysis to derive
the two-dimensional TDGL equation and study the shear band formation in the granular
shear flow.

At first, we add a small deviation to the most unstable mode as
$\mathbf{q}(\tau)=\mathbf{q}_c+\delta\mathbf{q}(\tau)$ and assume $\hat{\phi}_\mathrm{n}$
is unchanged if the deviation $\delta\mathbf{q}(\tau)$ is small
\begin{equation}
\hat{\phi}_\mathrm{n} \simeq A^{\rm L}(\xi,\zeta,\tau)\phi^{\rm L}_{q_c}
e^{i \mathbf{q}(\tau)\cdot\mathbf{z}} + \mathrm{c.c.}~, \label{eq:neutral-d}
\end{equation}
where we introduced $\mathbf{z}\equiv(\xi,\zeta)$ and the amplitude
$A^{\rm L}(\xi,\zeta,\tau)$ also depends on $\xi$. Combining the contribution from
the linearly stable mode $q_\xi\neq 0$ with $\hat{\phi}_\mathrm{n}$, we introduce the
\textit{hybrid solution}
\begin{eqnarray}
\hat{\phi}_\mathrm{h} &=& \left\{ A^{\rm L}(\xi,\zeta,\tau)\phi^{\rm L}_{q_c}
+ A^{\rm NL}(\xi,\zeta,\tau) \phi^{\rm NL}_{\mathbf{q}(\tau)} \right\}
e^{i \mathbf{q}(\tau) \cdot \mathbf{z}} + \mathrm{c.c.} \nonumber\\
&\simeq & A(\xi,\zeta,\tau)\left\{ \phi^{\rm L}_{q_c} +
\phi^{\rm NL}_{\mathbf{q}(\tau)} \right\}e^{i \mathbf{q}(\tau) \cdot \mathbf{z}}
+ \mathrm{c.c.}~, \label{eq:neutral2}
\end{eqnarray}
where $A^{\rm NL}(\xi,\zeta,\tau) $ and $\phi^{\rm NL}_{\mathbf{q}(\tau)}$ are the
amplitude and the Fourier coefficient, respectively, and we have used a strong
assumption that $A^{\rm L}(\xi,\zeta,\tau)$ and $A^{\rm NL}(\xi,\zeta,\tau)$ are
scaled by the common amplitude $A(\xi,\zeta,\tau)$. Because any modes $\mathbf{q}(\tau)$
with $q_\xi\neq 0$ are linearly stable, $\phi^{\rm NL}_{\mathbf{q}(\tau)}$ decays to
zero in the long time limit \cite{linear1,linear2,layer1,layer2,layer3,layer4,layer5}.

If we carry out the weakly nonlinear analysis by expanding $A(\xi,\zeta,\tau)$ as
\begin{equation}
A(\xi,\zeta,\tau) = \epsilon A_1(\xi,\zeta,\tau) + \epsilon^2 A_2(\xi,\zeta,\tau)
+ \epsilon^3 A_3(\xi,\zeta,\tau) + \dots~,
\end{equation}
and using $\hat{\phi}_\mathrm{h}$ instead of $\hat{\phi}_\mathrm{n}$, we find the
two-dimensional TDGL equation
\begin{equation}
\partial_\tau \bar{A}_1 = \sigma_c \bar{A}_1 + \bar{D}_1(\tau) \partial_\xi^2 \bar{A}_1
+ \bar{D}_2(\tau) \partial_\xi \partial_\zeta \bar{A}_1
+ \bar{D} \partial_\zeta^2 \bar{A}_1 + \bar{\beta} \bar{A}_1 |\bar{A}_1|^2
\label{eq:hybrid_GL3_red}
\end{equation}
of the rescaled amplitude $\bar{A}_1(\xi,\zeta,\tau)\equiv\epsilon^{1/2} A_1(\xi,\zeta,\tau)$
at $O(\epsilon^3)$, where $\bar{D}_1(\tau)$ and $\bar{D}_2(\tau)$ are the time dependent
diffusion coefficients (given by Eqs. (64) and (65) in Ref. \cite{saitoh1}). Similarly,
we also find the higher order equation
\begin{equation}
\partial_\tau \check{A} =
\sigma_c \check{A} + \bar{D}_1(\tau) \partial_\xi^2 \check{A} +
\bar{D}_2(\tau) \partial_\xi \partial_\zeta \check{A}
+ \bar{D} \partial_\zeta^2 \check{A} + \bar{\beta} \check{A} |\check{A}|^2 +
\epsilon \bar{\gamma} \check{A} |\check{A}|^4 + O(\epsilon^3) 
\label{eq:hybrid_GL5_red}
\end{equation}
of $\check{A}(\xi,\zeta,\tau)\equiv\epsilon^{1/2}\{A_1(\xi,\zeta,\tau)+\epsilon A_2(\xi,\zeta,\tau)+\epsilon^2 A_3(\xi,\zeta,\tau)\}$.

The two-dimensional TDGL equations (\ref{eq:hybrid_GL3_red}) and (\ref{eq:hybrid_GL5_red})
can be solved in the cases of the supercritical bifurcation and the subcritical bifurcation,
respectively. Because the time dependent diffusion coefficients $\bar{D}_1(\tau)$ and
$\bar{D}_2(\tau)$ are the functions of $\phi^{\rm NL}_{\mathbf{q}(\tau)}$, they decay to zero
as time goes on. Therefore, Eqs. (\ref{eq:hybrid_GL3_red}) and (\ref{eq:hybrid_GL5_red})
are respectively reduced to Eqs. (\ref{eq:red_TDGL3_layer}) and (\ref{eq:GL5_red}) in the
long time limit.

\section{Numerical solution of the TDGL equation}
In this section, we numerically solve the two-dimensional TDGL equations, where we find
the transient dynamics and the steady state of the solutions are qualitatively similar
to the evolution of the shear band in the two-dimensional granular shear flows.
\subsection{Numerical Method}
To solve Eqs. (\ref{eq:hybrid_GL3_red}) and (\ref{eq:hybrid_GL5_red}) numerically, we
prepare the $L^\ast\times L^\ast$ square box with the dimensionless system size $L^\ast\equiv L/d$
and divide the system into the $100\times 100$ grids. We discretize $\bar{A}_1(\xi,\zeta,\tau)$
and $\check{A}(\xi,\zeta,\tau)$ as $\bar{A}_{i,j}(\tau_k)$ and $\check{A}_{i,j}(\tau_k)$,
respectively, where the continuous variables are given by $\xi=i\times d\xi$,
$\zeta=j\times d\zeta$ and $\tau=k\times d\tau$ with the small increments $d\xi=d\zeta=L^\ast/100$
and $d\tau=1.0\times 10^{-4}$. We adopt the fourth order Runge-Kutta method to integrate
the time derivatives and the central difference method to calculate the diffusion terms,
e.g., the diffusion terms of $\bar{A}_{i,j}(\tau_k)$ are discretized as
\begin{eqnarray}
\partial_{\xi}^2 \bar{A}_{i,j}(\tau_k) &=&
\frac{\bar{A}_{i+1,j}(\tau_k)-2\bar{A}_{i,j}(\tau_k)+\bar{A}_{i-1,j}(\tau_k)}{d\xi^2}~,\\
\partial_{\xi}\partial_{\zeta} \bar{A}_{i,j}(\tau_k) &=&
\frac{\bar{A}_{i+1,j+1}(\tau_k)-\bar{A}_{i+1,j-1}(\tau_k)-\bar{A}_{i-1,j+1}(\tau_k)
+\bar{A}_{i-1,j-1}(\tau_k)}{4d\xi d\zeta}~,\\
\partial_{\zeta}^2 \bar{A}_{i,j}(\tau_k) &=&
\frac{\bar{A}_{i,j+1}(\tau_k)-2\bar{A}_{i,j}(\tau_k)+\bar{A}_{i,j-1}(\tau_k)}{d\zeta^2}~,
\label{eq:discretized-diffusion}
\end{eqnarray}
respectively. Since we adopt the Lees-Edwards boundary condition in the weakly nonlinear
analysis, we solve Eqs. (\ref{eq:hybrid_GL3_red}) and (\ref{eq:hybrid_GL5_red}) under
the periodic boundary conditions in the sheared frame. The initial values $\bar{A}_{i,j}(0)$
and $\check{A}_{i,j}(0)$ are given by the superpositions of the sine functions
$\sin(iK_\xi d\xi+jK_\zeta d\zeta)$ with the random wave numbers $K_\xi$ and $K_\zeta$.
\subsection{Results}
\begin{figure}
\includegraphics[width=18cm]{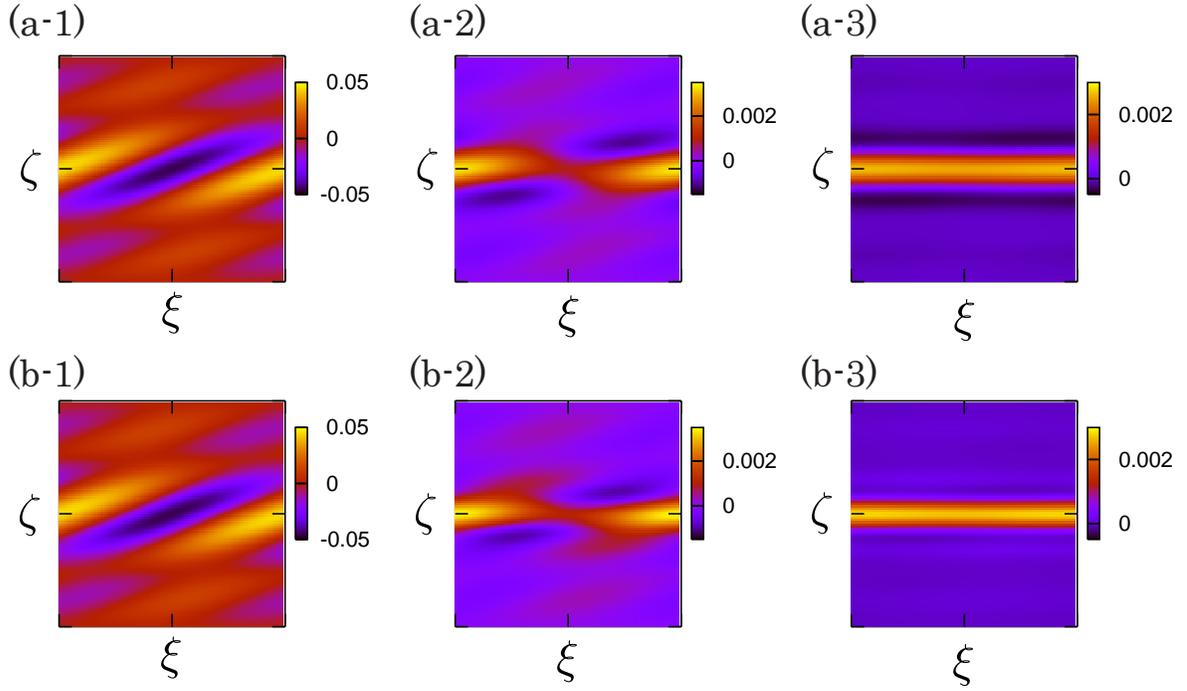}
\caption{Solutions of Eqs. (\ref{eq:hybrid_GL3_red}) and (\ref{eq:hybrid_GL5_red}),
where (a-1), (a-2) and (a-3) show the time evolution of $\bar{A}_1(\xi,\zeta,\tau)$
for the supercritical bifurcation ($\nu_0=0.20$), and (b-1), (b-2) and (b-3) show
the time evolution of $\check{A}(\xi,\zeta,\tau)$ for the subcritical bifurcation
($\nu_0=0.26$), respectively. Here, (a-1) and (b-1) are the results of $\tau=0.70$,
(a-2) and (b-2) are the results of $\tau=1.15$, and (a-3) and (b-3) are the results
of $\tau=1.35$, respectively.} \label{fig:ampl}
\end{figure}
\begin{figure}
\includegraphics[width=16cm]{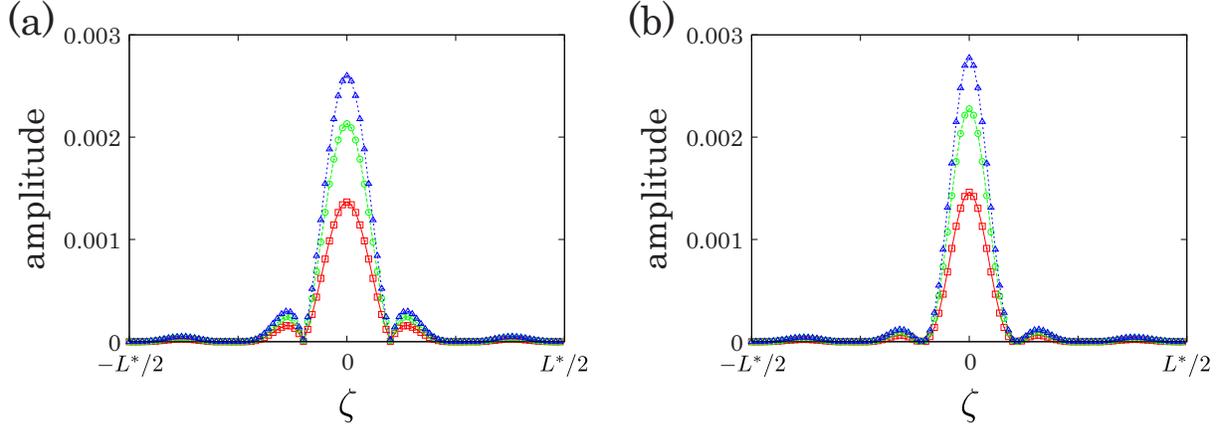}
\caption{Time evolution of the order parameter as a function of $\zeta$ for
(a) supercritical ($\nu_0=0.20$) and (b) subcritical ($\nu_0=0.26$) regimes, respectively.
The open squares, the open circles and the open triangles are the results of
$\tau=0.70$, $1.15$ and $1.35$, respectively.}
\label{fig:one}
\end{figure}
\begin{figure}
\includegraphics[width=8cm]{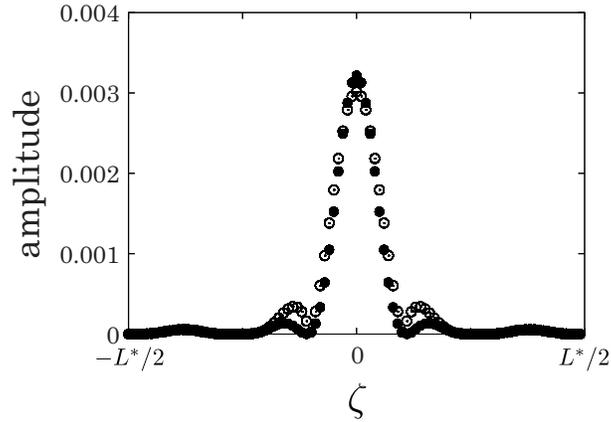}
\caption{Steady amplitudes as functions of $\zeta$. The open circles and the
closed circles represent $|\bar{A}_1|$ for $\nu_0=0.20$ and $|\check{A}|$ for
$\nu_0=0.26$, respectively.}
\label{fig:supsub}
\end{figure}
In our weakly nonlinear analysis, the coefficients $\bar{\beta}$ and $\bar{\gamma}$
are determined by the mean area fraction $\nu_0$ \cite{saitoh1}. If $\nu_0<0.245$,
$\bar{\beta}<0$ and the solution of Eq. (\ref{eq:hybrid_GL3_red}), i.e., $\bar{A}_1(\xi,\zeta,\tau)$,
converges. In this case, the supercritical bifurcation of the steady amplitude
is expected. If $0.245<\nu_0<0.275$, $\bar{\beta}>0$ and $\bar{\gamma}<0$. Thus, the
solution of Eq. (\ref{eq:hybrid_GL5_red}), i.e., $\check{A}(\xi,\zeta,\tau)$, converges and
the subcritical bifurcation of the steady amplitude is expected. Unfortunately,$\bar{\beta}>0$
and $\bar{\gamma}>0$ in the dense regime $\nu_0>0.275$ and neither Eqs. (\ref{eq:hybrid_GL3_red})
nor (\ref{eq:hybrid_GL5_red}) can be used. In the following, we use the small parameter
$\epsilon=0.01$ and show the numerical solutions of Eqs. (\ref{eq:hybrid_GL3_red})
and (\ref{eq:hybrid_GL5_red}) with $\nu_0=0.20$ and $0.26$, respectively.

Figure \ref{fig:ampl} displays the numerical solutions of the two-dimensional TDGL equations,
where (a-1), (a-2) and (a-3) are the time evolution of $\bar{A}_1(\xi,\zeta,\tau)$, and (b-1),
(b-2) and (b-3) are the time evolution of $\check{A}(\xi,\zeta,\tau)$, respectively.
In both cases, the disturbance in the short wave length is suppressed in the early stage
(Figs. \ref{fig:ampl}(a-1) and (b-1)) and the disturbance in the long wave length survives
(Figs. \ref{fig:ampl}(a-2) and (b-2)). Then, the shear band is generated in the center of
the system (Figs. \ref{fig:ampl}(a-3) and (b-3)). The steady amplitudes are homogeneous in
the $\xi$-direction, because the time dependent diffusion coefficients $\bar{D}_1(\tau)$
and $\bar{D}_2(\tau)$ disappear in the long time limit \cite{saitoh1}. As can be seen, we
cannot find any significant differences between $\bar{A}_1(\xi,\zeta,\tau)$ and $\check{A}(\xi,\zeta,\tau)$.

Figure \ref{fig:one} displays the numerical solutions averaged over the $\xi$-direction
in the supercritical (Fig. \ref{fig:one}(a)) and the subcritical (Fig. \ref{fig:one}(b))
regimes, respectively. Figure \ref{fig:supsub} displays the steady amplitudes, where the
shear bands have peaks at the center of the system $\zeta=0$. 

These results are qualitatively similar to the previous result of the area fraction
obtained by the DEM simulation \cite{saitoh}. The detailed comparison between the DEM
simulation and our analysis presented here will be reported elsewhere.
\section{Discussion and conclusion}
We numerically solved the two-dimensional TDGL equations (\ref{eq:hybrid_GL3_red}) and (\ref{eq:hybrid_GL5_red})
obtained by the hybrid approach to the weakly nonlinear analysis, where the structural evolution
and the steady state of the solutions are qualitatively similar to the shear band
observed in the DEM simulation. We also confirmed that the disturbance in the short wave
length is suppressed in the early stage, and the shear band is survived in the longest wave length.
The steady amplitudes are homogeneous in the sheared direction, which corresponds to the absence
of the time dependent diffusion coefficients $\bar{D}_1(\tau)$ and $\bar{D}_2(\tau)$ in the long
time limit. Neither the one-dimensional TDGL equation nor the Stuart-Landau equation
\cite{shukla1,shukla2,shukla3} cannot reproduce such a structural evolution of shear band.

In conclusion, the solutions of the two-dimensional TDGL equations reproduce the evolution of
shear band, which are similar to that observed in the DEM simulation \cite{saitoh}.

\begin{theacknowledgments}
This work was financially supported by an NWO-STW VICI grant.
Numerical computation in this work was carried out at the Yukawa Institute Computer Facility.
\end{theacknowledgments}



\bibliographystyle{aipproc}   
\bibliography{weakly}

\IfFileExists{weakly.bbl}{}
 {\typeout{}
  \typeout{******************************************}
  \typeout{** Please run "bibtex \jobname" to optain}
  \typeout{** the bibliography and then re-run LaTeX}
  \typeout{** twice to fix the references!}
  \typeout{******************************************}
  \typeout{}
 }
\end{document}